      [1999/12/01 v1.4c Il Nuovo Cimento]
\documentclass{cimento}
\usepackage{amsmath}
\usepackage{amsfonts}
\usepackage{amssymb}
\usepackage{dsfont}
\usepackage{latexsym}
\usepackage{epsfig}

\def\beq{\begin{equation}}
\def\eeq{\end{equation}}
\def\munu{{\mu\nu}}

\def\bitem{\begin{itemize}}
\def\eitem{\end{itemize}}
\def\d{\partial}

\def\di{\d_i}

\def\dn{\d_{\nu}}

\def\5M{\mathcal{M}^5}

\def\2f{{\phi}^2}
\def\teta{\vartheta}
\def\i{\hat{I}}
\def\j{\hat{J}}

\def\u{w}
\def\ds5{d s_{5}}
\def\bear{\begin{array}}
\def\ear{\end{array}}

\def\rmd{d}

\title{Hamiltonian Formulation of 5-dimensional Kaluza-Klein Theory}
\author{Valentino Lacquaniti\from{ICRA}\from{RomaTre}\ETC,
Giovanni Montani\from{ICRA}\from{ENEA}
}
\instlist{\inst{ICRA} ICRA, Physics Department (G9), Universit\`a  di Roma ``La Sapienza", Piazzale Aldo Moro 5, 00185 Rome, Italy. email: valentino.lacquaniti@icra.it, montani@icra.it
  \inst{RomaTre} Physics Department "E.Amaldi", Universit\'a di Roma "RomaTre",  Via della Vasca Navale 84, 00146 Rome, Italy. email: lacquaniti@fis.uniroma3.it
  \inst{ENEA} ENEA C.R. Frascati (Dipartimento F.P.N.), via Enrico Fermi 45, 00044
Frascati, Rome, Italy }
\PACSes{\PACSit{04.20.Fy}{04.50.+h}}
\begin{document}

\maketitle

\begin{abstract}
We analyze the consistency of the ADM approach to KK model; we prove that KK reduction commute with ADM splitting. This leads to a well defined Hamiltonian; we provide the outcome.  The electromagnetic constraint is derived from a geometrical one and this result enforces the physical meaning of KK model. Moreover we study the role of the extra scalar field $\phi$ we have in our model; classical hints from geodesic motion and cosmological solutions suggest that $\phi$ can be an alternative time variable in the relational point of view.
\end{abstract}

\section{Motivations}
The main argument of this work is to consider the ADM splitting of the 5D Kaluza-Klein (KK in the following) model.
The ADM splitting is usually requested when we want to consider the canonical approach to the dynamics of General Relativity(\cite{adm}); although we still have many unsolved problems it is today a very valuable approach towards quantum gravity, as shown for instance in Loop Quantum Gravity.
On the other hand, KK theories  provide interesting framework to unification theories ( \cite{kaluza},\cite{klein},\cite{acf}). The 5D KK model allows us to have a unified geometrical picture of gravitation with the elecromagnetic field. Starting from a 5D space-time, via the assumption of the cylindricity hypothesis and other restrictive conditions on our manifold, we are able to take into account the coupling of gravitation with U(1) gauge field plus an extra scala field $\phi$ (the scale factor governing the dynamics of the extra dimension). So far, we are interested in putting together these ideas in the same scheme. Our first task is to understand if ADM splitting is compatible with KK model. This question is not trivial, because in KK model we have to pay the price of breaking the 5D Poincar{\'e} simmetry; this could be not compatible with ADM splitting.  In next section we will show that we have two possible procedures to perform the ADM splitting of KK model; the dynamics we  have from these procedures could be not equivalent. Therefore this check can enforce or weaken the physical meaning of KK model itself. Moreover, if ADM splitting is available , we have as natural outcome the Hamiltonian formulation of the dynamics, useful for cosmological and quantum implications, expecially as far as the extra scalar field is concerned. There are some reasons to view this field as a time like degree of freedom, in the relational point of view. In the following we provide the proof of the compatibilty of the ADM slicing with the KK model and provide some classical hints about the  role of the scalar field as an alternative time-like variable. 
\section{ADM splitting of KK model}
In order to achieve the ADM slicing of the KK model we have two possible procedures: we label these as ADM-KK and KK-ADM procedures.
\subsection{KK-ADM}
In KK-ADM we start with the standard KK reduction of metric: starting by the 5D metric $J_{AB}$ we get the 4D metric $g_{\munu}$ , the U(1) gauge  4D vector $A_{\mu}$ and the scalar field $\phi$ in the usual way:
$$
\left\{
\bear{l}
J_{55}=-\phi^2\\
J_{5\mu}=-\phi^2A_{\mu} \\
J_{\munu}=g_{\munu}-\phi^2A_{\mu}A_{\nu}
\ear
\right.
$$
Then we can forget the extra dimension and split the 4D metric and the 4D gauge vector according to the $(3+1)$ ADM rules. We finally have: 
$$
J_{AB}  \Rightarrow \left\{
\bear{l}
g_{\munu} \rightarrow  \teta_{ij},S_i,N \\
A_{\mu}  \rightarrow  A_i,A_0 \\
\phi \rightarrow   \phi \\
\ear
\right.
\rightarrow
\left(
\begin{array}{ccc}
 N^2-S_iS^i-\2f A_0^2   &\quad  -S_i-\2f A_0A_i      & \quad -\2f A_0 \\
 -S_i^2\2f A_0A_i         &\quad -\teta_{ij}-\2fA_iA_j  &\quad  -\2f A_i \\
-\2f A_0                    & \quad -\2f A_i                   &\quad  -\2f
\end{array}
\right)
$$
Here $N,S_i,\teta_{ij}$ are the \textit{Lapse} function, the 3D \textit{Shift} vector, the 3D spatial induced metric; $A_0$ is the time component of the gauge vector and $A_i$ its spatial part $(A,B=0,1,2,3,5; \mu,\nu=0,1,2,3; i,j=1,2,3)$.
Although this way seems to be the easiest, in this procedure we have a not complete space-time slicing; indeed we are addressing a $3+1$ splitting in a 5D manifold, and in our construction the extra dimension is not included in the spatial separation. This is confirmed, as we will see, by the presence of second time derivatives in the final Lagrangian .
\subsection{ADM-KK}
In ADM-KK we perform as first the ADM splitting of the entire 5D model; in this way we deal with a $(4+1)$ ADM splitting and now the spatial extra dimension is included. Starting by the 5D metric we get the \textit{Lapse} function $N$, the 4D \textit{Shift} vector and the 4D spatial metrics $h_{\i\j}$. After this we can implement the KK reduction on $h_{\i\j}$ and we get a 3D gauge vector, a 3D spatial metric $\teta_{ij}$ and the extra scalar field $\phi$. 
$$
J_{AB}  \Rightarrow  \left\{
\bear{l}
h_{\i,\j} \, \rightarrow  A_i,\teta_{ij},\phi \\
N_{\i} \,\,\, \rightarrow  N_i, N_5 \\
N \,\,\,\, \rightarrow  N 
\ear
\right.
\Rightarrow
\left(
\begin{array}{ccc}
N^2-h_{\i\j}N^{\i}N^{\j} &\quad -N_{i} &\quad -N_5 \\
-N_{i} &\quad -\teta_{ij}-^2{}\2f  A_iA_j &\quad\ -\2f A_i \\
-N_5 &\quad -\2f A_{i} &\quad -\2f
\end{array}
\right)
$$
Here $N$ and $N_{\i}$ are the \textit{Lapse} and the 4D \textit{Shift} vector ($\i,\j=1,2,3,5$).
Now, by construction, our slicing is complete but, as we can see, the set of variables lacks explicitely the component $A_0$.
Therefore, at this step, our procedures seem to lead to different outcomes; we have two different metrics  with two set of unrelated variables. Then we don't know if our procedures commute and we are not able to have a unique well-defined Hamiltonian. Moreover, even if we would take one of them as viable, nevertheless, at the present stage each of them looks unsatisfactory. 
However, we are dealing in both procedures with objects that must show well defined properties of transformations under the same class of pure spatial KK diffeomorphisms. At this step the transformation laws of $N_{\i}$ are still ambiguous, while for the others variables we have clearly 3D tensorial laws or gauge laws. This allows us to look for "`conversion formulas "` betweeen the two metrics; this analysys consists in the following steps: we implement the KK restrictive hypothesis on the Shift functions $N_{\i}$; by this we recognize that they are not a pure spatial vector 
neither a simple gauge vector but a mixture of them; we study in details the laws of transformations by the tetradic picture and finally  perform a "`positional"' comparison between the two metrics.
At the end we are able to state these "`conversion formulas"'(\cite{lm}); by these formulas we can recognize  equivalence for all components of the metrics an the inverse metrics.
$$
\left\{
\begin{array}{l}
N_{i}=S_{i}+ \phi^2A_0A_{i} \\
N_5=\2fA_0
\end{array}
\right.
\quad\quad\quad
\left\{
\begin{array}{l}
N^{i}=S^{i} \\
N^5=N^2A^0
\end{array}
\right.
$$
The real physical meaning of these formulas is recognized once we compare the Lagrangians and calculate the Hamiltonian.
By these formulas , we can recast the two Lagrangians that arise from our procedures in the same set of variables. We choose the set that contains $A_0$ and we have:
$$
L_{adm-kk}=L_0 +\frac{1}{8\pi G} \sqrt{\teta}NK\d_{\eta}\phi \quad\quad
L_{kk-adm}=L_0+\frac{1}{8\pi G} \sqrt{\teta}N\d_{\eta}\d_{\eta}\phi
$$
Here $K$ is the trace of the 3D extrinsic curvature and we have $\d_{\eta}=(1/N)(\d_0-S^i\d_i)$.
$L_0$ is a term, egual for both Lagrangians, that contains no derivatives of $\phi$.  The additives terms with derivatives of $\phi$ are found to be equivalent apart from  surface terms. Hence, if boundary conditions allow us to discard these terms  we can conlude that we are dealing with equivalent dynamics and that our conversions formulas have a real physical meaning.
Thus we  conclude that ADM splitting and KK reduction commute; we are able in both cases to build a complete space-time slicing without missing the time component of the gauge vector: this is a positive check for KK model.
Moreover, due to the commutation we can have a unique Hamiltonian; we provide the result(\cite{lm}):
$$
\mathcal{H}=NH^{N}+S_iH^{i}+A_{0}H^0
$$
$$
\left\{
\begin{array}{l}
H^{N}= b\sqrt{\teta}(\phi R-2D^i\di\phi-\frac{1}{4}\phi^3F_{ij}F^{ij})-\frac{1}{2b\sqrt{\teta}\phi}T_{ijkl}\Sigma^{ij}\Sigma^{kl}-\frac{1}{6b\sqrt{\teta}}\pi^2_{\phi}\phi + \\ 
\quad\quad\ \
+\frac{1}{3b\sqrt{\teta}}\pi_{\phi}\Sigma^{ij}\teta_{ij} 
 -\frac{2}{b\sqrt{\teta}\phi^3}\pi^i\pi^j\teta_{ij} \\
H^i= -2D_j\Sigma^{ij}+\pi_{\phi}\d^{i}\phi-\pi^jF^{i\ }_{\ j} \\
H^0=-D_i\pi^i \\
\end{array}
\right.
$$
Here $b=-1/16\pi G$; $\Sigma^{ij},\pi^i,\pi_{\phi}$ are respectively the conjugates momenta to $\teta_{ij},A_i,\phi$. $F_{ij}$ is the spatial part of the electromagnetic tensor, $D_i$ is the 3D covariant derivative and the supermetric $T_{ijkl}$ reads  as follows: $T_{ijkl}=(-\frac{2}{3}\teta_{ij}\teta_{kl}+\teta_{ik}\teta_{jl}+\teta_{il}\teta_{jk})$.
We rebuild  the electromagnetic constraint: $A_0$ is a Lagrangian multiplier; this is   predicted by the conversion formulas ( $A_0 \infty N_5 $).  After the KK reduction one of our geometrical constraints becomes a gauge constraints. Therefore the proof of the commutation enforces the physical meaning of KK model, ensures us to have a unique Hamiltonian and gives insight in the understanding of gauge symmetry generation . 
Finally it can be a first step in the Ashtekar reformulation (\cite{abbay})of KK model.

\section{The role of the scalar field}

In our Hamiltonian we have the presence of the scalar field and its conjugate momentum, with a non trivial dynamics (
for instance,the presence of $\phi$ affects the supermetric, as we notice by seeing the factor $2/3$ in the supermetrics  ). There are some reasons to think that $\phi$ can be an alternative time variable in the relational point of view and hence solve the problem of the frozen formalism. The advantage of KK model in this is that the scalar field is naturally provided by the theory and we don't need to consider matter or to put extra field by hands.  This kind of study is still "`in progress "`. 
In the following we consider two classical application as hints  for the interpretation of the scalar field as time variable.

\subsection{5D Geodesic motion} 
Equations for geodesic motions arise from the variation of the action $S=\int \ds5$, where the 5D line element reads:
$
\ds5=g_{\munu}\rmd x^{\mu}\rmd x^{\nu}-\2f (A_{\mu}\rmd x^{\mu}+\rmd x^5)^2
$
We vary this action and after recast the equation in term of $ds$ rather than $\ds5$; we get:
$$
\left\{
\bear{l}
\frac{d}{ds}\u_5\equiv0 \\
\frac{D}{Ds}u^{\mu}=F^{\munu}u_{\nu}(\frac{\u_5}{\alpha})+\frac{1}{2\phi^4}(u^{\mu}u^{\nu}-g^{\munu})\dn\2f(\frac{\u_5^2}{\alpha^2})
\ear
\right.
\label{sistema1}
$$
Here
$w_5$ is the fifth covariant component of the 5D velocity ($w^A=\frac{dx^A}{ds_5}$), $u^{\mu}$ is the usual 4D velocity ($u^{\mu}=\frac{dx^{\mu}}{ds}$) and $\frac{D}{Ds}$ is the usual covariant derivative along the curve. As we can see $w_5$ is a constant of motion ; it explicetely reads $w_5=-\phi^2(A_{\mu}\u^{\mu}+w^5)$ and moreover, is a scalar under KK transformations. Finally,
$\alpha$ reads as follows:
$
\alpha=\frac{\rmd s}{\ds5}=\sqrt{1+\frac{\u^2_5}{\phi^2}}
$
Hence, we can define the charge-mass ratio for a test particle in terms of kinematical objects. By adopting proper dimensional constant we have:
$$
q/\sqrt{4G}m= w_5(1+\frac{w_5^2}{\phi^2})^{-1}
$$
Due to presence of  $\phi$ however, the charge-mass ratio is not conserved. But, if we assume $\phi >>1 $ (i.e. $\phi>10^{21}$ for the electron) we can recover the usual electrodynamics; in this case indeed
the conservation of $q/m$ is restored at a satisfactory degree of approximation; the dynamical term in $\phi$ is negligible and we can have realistic value for the charge-mass ratio (no planckian mass values).
Anyway  a time varying charge is very interesting and this is indeed the relevant feature of this naive model;
an isotropic, slow time-varying $\phi$ can explain the variation in time of the fine structure constant over cosmological scale (which seems to be inferred by recent analysis (\cite{alfa}). Hence, by this point of view $\phi$ can be viewed as a mark of the evolution of the universe.

\subsection{Cosmological Solutions}
{\emph{Friedmann-like solution: }}
 in this simple model we require flatness, homogenity, isotropy of space, no electromagnetic field. The Lagrangian written in ADM variables simplifys a lot; we deal only with $\phi$ and with the 3D scale factor $a(t)$ and finally we have: 
$$
\left\{
\bear{l}
\frac{\ddot{a}}{a}+(\frac{\dot{a}}{a})^2=0 \\
\frac{{\d}^2}{\d t^2} (\phi a)=0
\ear
\right.
\Rightarrow
\left\{
\bear{l}
a(t)=a_0\,t^{\frac{1}{2}} \\
\phi=\beta't^{\frac{1}{2}}+\gamma'\frac{1}{t^{\frac{1}{2}}}
\ear
\right.
$$
We have a power-law behaviour, and we can recognize a Kasner solution for an epanding/collapsing extra-dimension.
{\emph{DeSitter-like solution:}}
we add a positive Cosmological Constant $\Lambda$ to the previous model and we get:
$$
\left\{
\bear{l}
\frac{\ddot{a}}{a}+(\frac{\dot{a}}{a})^2=\Lambda \\
\frac{{\d}^2}{\d t^2} (\phi a)=\phi a\Lambda 
\ear
\right.
\Rightarrow
\left\{
\bear{l}
a(t)=a_0 e^{\sqrt{\frac{\Lambda}{2}}\,t} \\
\phi_{-}=\beta' \frac{ e^{-\sqrt{\Lambda}\,t}}{e^{\sqrt{\frac{\Lambda}{2}}\,t}} \\
\phi_{+}=\beta' \frac{ e^{\sqrt{\Lambda}\,t}}{e^{\sqrt{\frac{\Lambda}{2}}\,t}}
\ear
\right.
$$
These are only simple solutions ( and, moreover, in vacuum ), but they show a strongly link between the scale factor and the scalar field: in both cases the scalar field mimics the behaviour of the scale factor.

\section{Final Remarks}
The ADM approach to 5D KK model is consistent and  ADM splitting and KK reduction commute; we have a physical interpretation of this feature ( $A_0 \infty N_5 $), that gives insight in the understanding of gauge summetry generation. There are reasons to think that ADM splitting also works in non-abelian KK models and it seems an interesting check for a future work. This result is also relevant for the selfconsistency of KK model itself and a for a well defined Hamiltonian description that could take into account the 5D Poincar{\'e} symmetry breaking scenario.
The scalar field  shows some classical feature ( in simple models) that we can link to its possible role  as a time-like variable in the relational point of view. Hence, in order to deep understand the role of $\phi$ is worth dealing  with  matter fields  and with a detailed cosmological solution  and, eventually, with the reformulation of the model in terms of Ashtekar variables.

\end{document}